\documentclass[twocolumn,showpacs,preprintnumbers,amsmath,amssymb,aps,pra]{revtex4}

\usepackage{graphicx}
\usepackage{dcolumn}
\usepackage{bm}
\usepackage{color}
\usepackage{times}
\usepackage{wasysym}
\renewcommand{\vec}[1]{\mathbf{#1}}
\usepackage{graphicx}
\usepackage{times}

\newcommand{\ch}{\ensuremath{\chi^{(3)}\text{ }}} 
\newcommand{\vE}[1]{\ensuremath{\mathbf{E}_{#1}}}
\newcommand{\vEc}[1]{\ensuremath{\mathbf{E}^*_{#1}}}
\newcommand{\vx}{\ensuremath{\mathbf{x}}}
\newcommand{\w}{\ensuremath{\omega}}
\newcommand{\e}{\ensuremath{\epsilon}}
\newcommand{\lp}{\ensuremath{\left(}}
\newcommand{\rp}{\ensuremath{\right)}}
\newcommand{\ws}[1]{\ensuremath{\omega_{#1}}}
\renewcommand{\a}{\ensuremath{\alpha}}
\newcommand{\wpci}{\ensuremath{\omega_\text{cp}^{\text{crit}}~}}
\newcommand{\wunit}{\ensuremath{\left({2\pi c\over a} \right)}}

\newcommand{\punit}{\ensuremath{\left({2\pi c\epsilon_0 a^2\over\chi^{(3)}}\right)}}
\newcommand{\ablunit}{\ensuremath{\left({\chi^{(3)}\over\epsilon_0 a^2 h}\right)}}
\newcommand{\abunit}{\ensuremath{\left({\chi^{(3)}\over\epsilon_0 a^3}\right)}}
\newcommand{\pcrit}{\ensuremath{P_0^{\text{crit}}~}}

\renewcommand{\eqref}[1]{Eq.~\ref{eq:#1}}
\newcommand{\eqreftwo}[2]{Eqs.~\ref{eq:#1}--\ref{eq:#2}}

\newcommand{\figref}[1]{Fig.~\ref{fig:#1}}
\newcommand{\Figref}[1]{Figure~\ref{fig:#1}}
\newcommand{\figreftwo}[2]{Figs.~\ref{fig:#1} and~\ref{fig:#2}}

\newcommand{\secref}[1]{Sec.~\ref{sec:#1}}

\newcommand{\citeasnoun}[1]{Ref.~\onlinecite{#1}}

\begin{document}

\title{High-efficiency degenerate four wave-mixing in triply resonant nanobeam cavities}

\author{Zin Lin$^1$}
\email{zlin@seas.harvard.edu}
\author{Thomas Alcorn$^2$}
\author{Marko Loncar$^1$}
\author{Steven G. Johnson$^2$}
\author{Alejandro W. Rodriguez$^{3}$}%
\affiliation{$^1$School of Engineering and Applied Sciences, Harvard University, Cambridge, MA 02138}
\affiliation{$^2$Department of Mathematics, Massachusetts Institute of Technology, Cambridge, MA 02139}
\affiliation{$^3$Department of Electrical Engineering, Princeton University, Princeton, NJ, 08544}

\date{\today}

\begin{abstract}
  We demonstrate high-efficiency, degenerate four-wave mixing in
  triply resonant Kerr ($\ch$) photonic crystal (PhC) nanobeam
  cavities. Using a combination of temporal coupled mode theory and
  nonlinear finite-difference time-domain (FDTD) simulations, we study
  the nonlinear dynamics of resonant four-wave mixing processes and
  demonstrate the possibility of observing high-efficiency limit
  cycles and steady-state conversion corresponding to $\approx 100\%$
  depletion of the pump light at low powers, even including effects
  due to losses, self- and cross-phase modulation, and imperfect
  frequency matching. Assuming operation in the telecom range, we
  predict close to perfect quantum efficiencies at reasonably low
  $\sim 50\mathrm{m}W$ input powers in silicon micrometer-scale
  cavities.
\end{abstract}

\pacs{Valid PACS appear here}%
\maketitle

\section{Introduction}

Optical nonlinearities play an important role in numerous photonic
applications, including frequency conversion and
modulation~\cite{Boyd92,Hald01,Lifshitz05,Morozov05,yeh07,Hebling04,Ruan09},
light amplification and
lasing~\cite{Boyd92,Stolen82,Kroll62,Stolen72}, beam
focusing~\cite{Boyd92,Akhmanov66}, phase
conjugation~\cite{Boyd92,Fisher83}, signal
processing~\cite{Contestabile04,Yoo96}, and optical
isolation~\cite{Gallo01,Lira12}.  Recent developments in
nanofabrication are enabling fabrication of nanophotonic structures,
e.g. waveguides and cavities, that confine light over long times and
small volumes~\cite{Asano03,Almeida04,Vlasov05,Song05,Deotare09},
minimizing the power requirements of nonlinear
devices~\cite{Kippenberg04,Rodriguez07:OE} and paving the way for
novel on-chip applications based on all-optical nonlinear
effects~\cite{Almeida04,Soljacic02:bistable,Ilchenko04,Furst10,Liu10,Foster06,Bieler08,Hamam08,Bermel07,Bravo07:rev,Caspani11}. In
addition to greatly enhancing light--matter interactions, the use of
cavities can also lead to qualitatively rich dynamical phenomena,
including multistability and limit
cycles~\cite{Felber76,Dumeige11,Smith70,Hashemi09,Drummond80,Grygiel92,Abraham82}. In
this paper, we explore realistic microcavity designs that enable
highly efficient degenerate four-wave mixing (DFWM) beyond the
undepleted pump regime. In particular, we extend the results of our
previous work~\cite{Ramirez11}, which focused on the theoretical
description of DFWM in triply resonant systems via the temporal
coupled-mode theory (TCMT) framework, to account for various realistic
and important effects, including linear losses, self- and cross-phase
modulation, and frequency mismatch. Specifically, we consider the
nonlinear process depicted in \figref{drawing}, in which incident
light at two nearby frequencies, a pump $\omega_0$ and signal
$\omega_\text{m} = \ws{0}-\Delta\omega$ photon, is up-converted into
output light at another nearby frequency, an idler $\omega_\text{p} =
\omega_0 + \Delta\omega$ photon, inside a triply resonant photonic
crystal nanobeam cavity (depicted schematically in
~\figref{3Dfields}). We demonstrate that 100\% conversion efficiency
(complete depletion of the pump power) can be achieved at a critical
power and that detrimental effects associated with self- and
cross-phase modulation can be overcome by appropriate tuning of the
cavity resonances. Surprisingly, we find that critical solutions
associated with maximal frequency conversion are ultra-sensitive to
frequency mismatch (deviations from perfect frequency matching
resulting from fabrication imperfections), but that there exist other
robust, dynamical states (e.g. ``depleted'' states and limit cycles)
that, when properly excited, can result in high conversion
efficiencies at reasonable pump powers. We demonstrate realistic
designs based on PhC nanobeam cavities that yield 100\% conversion
efficiencies at $\sim 50\mathrm{m}W$ pump powers and over broad
bandwidths (modal lifetimes $Q \sim 1000$s). Although our cavity
designs and power requirements are obtained using the TCMT framework,
we validate these predictions by checking them against rigorous,
nonlinear FDTD simulations.

Although chip-scale nonlinear frequency conversion has been a topic of
interest for decades~\cite{Caspani11}, most theoretical and
experimental works have been primarily focused on large-etalon and
singly resonant systems exhibiting either large footprints and small
bandwidths~\cite{Ilchenko04,Furst10,Krishnan02,Kuo11}, or low
conversion efficiencies (the undepleted pump
regime)~\cite{Kippenberg04,Ferrera08,Absil00,Dumeige06}. These include
studies of $\chi^{(2)}$ processes such as second harmonic generation
\cite{levy11,rivoire11b:apl,Furst10,Buckley13}, sum and difference
frequency generation \cite{Rivoire10}, and optical parametric
amplification~\cite{Foster06,Liu10,Kuyken11}, as well as \ch processes
such as third harmonic generation~\cite{Carmon07,levy11}, four-wave
mixing~\cite{Fukuda05, Reza08,Agha12} and optical parametric
oscillators~\cite{Kippenberg04,Kippenberg07,Levy10,Okawachi11}. Studies
that go beyond the undepleted regime and/or employ resonant cavities
reveal complex nonlinear dynamics in addition to high efficiency
conversion~\cite{Drummond80,Rodriguez07:OE,Hashemi09,Grygiel92,Burgess09:OE,Ramirez11,Zhuan12},
but have primarily focused on ring resonator geometries due to their
simplicity and high degree of tunability~\cite{Zhuan12}. Significant
efforts are underway to explore similar functionality in
wavelength-scale photonic components (e.g. photonic crystal
cavities)~\cite{Rivoire10,Buckley13}, although high-efficiency
conversion has yet to be experimentally demonstrated.  Photonic
crystal nanobeam cavities not only offer a high degree of tunability,
but also mitigates the well-known volume and bandwidth tradeoffs
associated with ring resonators~\cite{Joannopoulos95}, yielding
minimal device footprint and on-chip
integrability~\cite{Sauvan05,Zain08}, in addition to high quality
factors~\cite{Murray08,Notomi08,Deotare09,Yinan09,Qimin11}.

In what follows, we investigate the conditions and design criteria
needed to achieve high efficiency DFWM in realistic nanobeam
cavities. Our paper is divided into two primary sections. In
\secref{TCMT}, we revisit the TCMT framework introduced
in~\citeasnoun{Ramirez11}, and extend it to include new effects
arising from cavity losses (\secref{loss}), self- and cross-phase
modulation (\secref{SPM-XPM}), and frequency mismatch
(\secref{mismatch}). In \secref{designs}, we consider specific
designs, starting with a simple 2d design (\secref{2d}) and concluding
with a more realistic 3d design suitable for experimental realization
(\secref{3d}). The predictions of our TCMT are checked and validated
in the 2d case against exact nonlinear FDTD simulations.

\section{Temporal coupled-mode theory}
\label{sec:TCMT}

\begin{figure}[t!]
 \centering
 \includegraphics[scale=0.25]{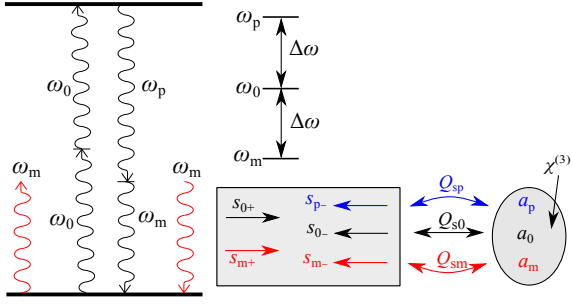}
 \caption{Schematic diagram of a degenerate four-wave mixing process
   in which a pump photon at frequency $\ws{0}$ and a signal photon at
   frequency $\ws{\text{m}} = \ws{0} - \Delta \w$ are converted into an idler
   photon at $\ws{\text{p}} = \w_0 + \Delta \w$ and an additional signal
   photon at $\w_\text{m}$, inside of a triply resonant $\chi^{(3)}$
   nonlinear cavity. The cavity supports three resonant modes with
   frequencies $\omega_{\text{c}k}$, lifetimes $Q_{k}$, and modal amplitudes
   $a_k$, which are coupled to a waveguide supporting propagating
   modes at the incident/output frequencies $\omega_k$, with coupling
   lifetimes $Q_{\text{s}k}$. The incident and output powers associated with
   the $k$th mode are given by $|s_{k+}|^2$ and $|s_{k-}|^2$.}
 \label{fig:drawing}
\end{figure}

In order to obtain accurate predictions for realistic designs, we
extend the TCMT predictions of ~\cite{Ramirez11} to include important
effects associated with the presence of losses, self- and cross-phase
modulation, and imperfect frequency-matching. We consider the DFWM
process depicted in \figref{drawing}, in which incident light from
some input/output channel (e.g. a waveguide) at frequencies $\omega_0$
and $\omega_\text{m}$ is converted to output light at a different frequency
$\omega_\text{p} = 2 \omega_0 - \omega_\text{m}$ inside a triply-resonant $\ch$
cavity. The fundamental assumption of TCMT (accurate for weak
nonlinearities) is that any such system, regardless of geometry, can
be accurately described by a few set of geometry-specific
parameters~\cite{Ramirez11}. These include, the frequencies
$\omega_{\text{c}k}$ and corresponding lifetimes $\tau_k$ (or quality factors
$Q_k = \w_{\text{c}k} \tau_k/2$) of the cavity modes, as well as nonlinear
coupling coefficients, $\alpha_{kk'}$ and $\beta_{k}$, determined by
overlap integrals between the cavity modes (and often derived from
perturbation theory~\cite{Rodriguez07:OE}). Note that in general, the
total decay rate ($1/\tau_k$) of the modes consist of decay into the
input/output channel ($1/\tau_{\text{s}k}$), as well as external
(e.g. absorption or radiation) losses with decay rate $1/\tau_{\text{e}k}$,
so that $1/\tau_k = 1/\tau_{\text{s}k} + 1/\tau_{\text{e}k}$. Letting $a_k$ denote
the time-dependent complex amplitude of the $k$th cavity mode
(normalized so that $|a_k|^2 $ is the electromagnetic energy stored in
this mode), and letting $s_{k\pm}$ denote the time-dependent amplitude
of the incident (+) and outgoing ($-$) light (normalized so that
$|s_{k\pm}|^2$ is the power at the incident/output frequency
$\omega_k$), it follows that the field amplitudes are determined by
the following set of coupled ordinary differential
equations~\cite{Rodriguez07:OE}:
\begin{align}
{da_0 \over dt} &= i \omega_\text{c0} \left( 1 - \alpha_{00} |a_0|^2 - \alpha_\text{0m} |a_\text{m}|^2 - \alpha_\text{0p} |a_\text{p}|^2 \right) a_0  \notag \\
		 &- {a_0 \over \tau_0} - i \omega_\text{c0} \beta_0 a_0^* a_\text{m} a_\text{p} + \sqrt{{2 \over \tau_\text{s0}}} s_{0+}, \label{eq:cme1}\\
{da_\text{m} \over dt} &= i \omega_\text{cm} \left( 1 - \alpha_\text{m0} |a_0|^2 - \alpha_\text{mm} |a_\text{m}|^2 - \alpha_\text{mp} |a_\text{p}|^2 \right) a_\text{m}  \notag \\
		 &- {a_\text{m} \over \tau_\text{m}} - i \omega_\text{cm} \beta_\text{m} a_0^2 a_\text{p}^* + \sqrt{{2 \over \tau_\text{sm}}} s_{\text{m}+}, \label{eq:cme2} \\
{da_\text{p} \over dt} &= i \omega_\text{cp} \left( 1 - \alpha_\text{p0} |a_0|^2 - \alpha_\text{pm} |a_\text{m}|^2 - \alpha_\text{pp} |a_\text{p}|^2 \right) a_\text{p}  \notag \\
		 &- {a_\text{p} \over \tau_\text{p}} - i \omega_\text{cp} \beta_\text{p} a_0^2 a_\text{m}^*, \label{eq:cme3}\\
s_{0-} &= \sqrt{{2 \over \tau_\text{s0}}} a_0 - s_{0+}, \quad s_{\text{m}-} = \sqrt{{2 \over \tau_\text{sm}}} a_\text{m} - s_{\text{m}+}, \notag \\
s_{\text{p}-} &= \sqrt{{2 \over \tau_\text{sp}}} a_\text{p},
\label{eq:cme4}
\end{align}
where the nonlinear coupling coefficients~\cite{Ramirez11},
\small
\begin{align}
\alpha_{kk} &=  {1 \over 8} \frac{\int d^3\vx ~ \e_0 \ch \left[ 2 |\vE{k} \cdot \vEc{k}|^2 + |\vE{k} \cdot \vE{k}|^2 \right]}{\lp \int d^3 \vx ~ \epsilon |\vE{k}|^2 \rp^2} \label{eq:nlcoef2} \\
\alpha_{kk'} &=  {1 \over 4} \frac{\int d^3\vx ~ \e_0 \ch \left[ |\vE{k} \cdot \vEc{k'}|^2 + |\vE{k} \cdot \vE{k'}|^2 + |\vE{k}|^2 |\vE{k'}|^2 \right]}{\lp \int d^3 \vx ~ \epsilon |\vE{k}|^2 \rp \lp \int d^3 \vx ~ \epsilon |\vE{k'}|^2 \rp} \label{eq:nlcoef3}
\end{align}
\begin{equation}
\alpha_{kk'} = \alpha_{k'k} \label{eq:nlcoef4}
\end{equation}
\begin{equation}
  \beta_0 = {1 \over 4} \frac{\int d^3\mathbf{x} \epsilon_0 \chi^{(3)} \left[ ( \vEc{0} \cdot \vEc{0} ) (\vE{\text{m}} \cdot \vE{\text{p}}) +2 (\vEc{0} \cdot \vE{\text{m}})(\vEc{0} \cdot \vE{\text{p}}) \right]}{\lp \int d^3 \vx ~ \epsilon |\vE{0}|^2 \rp \lp \int d^3 \vx ~ \epsilon |\vE{\text{m}}|^2 \rp^{1/2} \lp \int d^3\vx ~ \epsilon |\vE{\text{p}}|^2 \rp ^{1/2}} \label{eq:beta}
\end{equation}
\begin{equation}
\beta_\text{m} = \beta_\text{p} = \beta^*_0 /2 \label{eq:nlcoef1}
\end{equation}
\normalsize express the strength of the nonlinearity for a given mode,
with the $\alpha$ terms describing SPM and XPM effects and the $\beta$
terms characterizing energy transfer between the modes. (Technically
speaking, this qualitative distinction between $\alpha$ and $\beta$
is only true in the limit of small losses~\cite{Rodriguez07:OE}).

\subsection{Losses}
\label{sec:loss}

\eqreftwo{cme1}{cme4} can be solved to study the steady-state conversion
efficiency of the system [$\eta = |s_{\text{p}-}|^2 / (|s_{0+}|^2 +
|s_{\text{m}+}|^2)$] in response to incident light at the resonant cavity
frequencies ($\omega_k = \omega_{\text{c}k}$), as was done in
\citeasnoun{Ramirez11} in the ideal case of perfect frequency-matching
($\omega_\text{cp} = 2\omega_\text{c0} - \omega_\text{cm}$), no losses ($\tau_k \neq
\tau_{\text{s}k}$), and no self- or cross-phase modulation ($\alpha=0$). In
this ideal case, one can obtain analytical expressions for the maximum
efficiency $\eta^\text{max}$ and critical powers, $P^\text{crit}_0 =
|s^\text{crit}_{0+}|^2$ and $P^\text{crit}_\text{m} =
|s_{\text{m}+}^\text{crit}|^2$, at which 100\% depletion of the total input
power is attained~\cite{Ramirez11}. Performing a similar calculation,
but this time including the possibility of losses, we find:
\begin{align}
\label{eq:critp}
P_0^{\text{crit}} &= \frac{4}{\tau_\text{s0} |\beta_0| \sqrt{\tau_\text{m} \tau_\text{p} \w_\text{m} \w_\text{p}}} \\
\eta^{\text{max}} &= {\tau_\text{p} \over \tau_\text{sp}} \lp 2 - {\tau_\text{s0} \over
  \tau_{0}} \rp {\w_\text{p} \over 2 \w_0}.
\label{eq:criteta} 
\end{align}
With respect to the lossless case, the presence of losses merely
decreases the maximum achievable efficiency by a factor of $\tau_\text{p} /
\tau_\text{sp} (2 - \tau_\text{s0}/\tau_0)$ while increasing the critical power
$P^\text{crit}_0$ by a factor of $\sqrt{\tau_\text{sm}\tau_\text{sp} / \tau_\text{m}
  \tau_\text{p}}$. As in the case of no losses, 100\% depletion is only
possible in the limit as $P_\text{m} \to 0$, from which it follows that the
maximum efficiency is independent of $\tau_\text{m}$. As noted
in~\cite{Ramirez11}, the existence of a limiting efficiency
(\eqref{criteta}) can also be predicted from the Manley--Rowe
relations governing energy transfer in nonlinear systems~\cite{Haus84}
as can the limiting condition $P_\text{m} \to 0$. While theoretically this
suggests that one should always employ as small a $P_\text{m}$ as possible,
as we show below, practical considerations make it desirable to work
at a small but finite (non-negligible) $P_\text{m}$.

\subsection{Self- and cross-phase modulation}
\label{sec:SPM-XPM}

Unlike losses, the presence of self- and cross-phase modulation
dramatically alters the frequency-conversion process. Specifically, a
finite $\alpha$ leads to a power-dependent shift in the effective
cavity frequencies $\omega^\text{NL}_{\text{c}k} = \omega_{\text{c}k}
(1 - \sum_j \alpha_{kj} |A_j|^2)$ that spoils both the
frequency-matching condition as well as the coupling of the incident
light to the corresponding cavity modes. One approach to overcome this
difficulty is to choose/design the linear cavity frequencies to have
frequency $\omega_{\text{c}k}$ slightly detuned from the incident
frequencies $\omega_k$, such that at the critical powers, the
effective cavity frequencies align with the incident frequencies and
satisfy the frequency matching
condition~\cite{Ramirez11}. Specifically, assuming incident light at
$\omega_0$ and $\omega_\text{m}$, it follows by inspection of
\eqreftwo{cme1}{cme4} that preshifting the linear cavity resonances
away from the incident frequencies according to the transformation,
\begin{align}
  \w_\text{c0}^{\text{crit}} &= \frac{\w_0} { 1 - \alpha_{00} |a_0^{\text{crit}}|^2 - \alpha_\text{0m} |a_\text{m}^{\text{crit}}|^2 - \alpha_\text{0p} |a_\text{p}^{\text{crit}}|^2 } \label{eq:preshift1} \\
  \w_\text{cm}^{\text{crit}} &= \frac{\w_\text{m}} { 1 - \alpha_\text{m0} |a_0^{\text{crit}}|^2 - \alpha_\text{mm} |a_\text{m}^{\text{crit}}|^2 - \alpha_\text{mp} |a_\text{p}^{\text{crit}}|^2 } \label{eq:preshift2} \\
  \w_\text{cp}^{\text{crit}} &= \frac{2\w_0 - \w_\text{m}} { 1 - \alpha_\text{p0}
    |a_0^{\text{crit}}|^2 - \alpha_\text{pm} |a_\text{m}^{\text{crit}}|^2 -
    \alpha_\text{pp} |a_\text{p}^{\text{crit}}|^2 }, \label{eq:preshift3}
\end{align}
yields the same steady-state critical solution obtained for
$\alpha=0$, where $a_k^\text{crit}$ denote the critical, steady-state
cavity fields. 

An alternative approach to excite the critical solution above in the
presence of self- and cross-phase modulation is to detune the incident
frequencies away from $\omega_\text{c0}$ and $\omega_\text{cm}$, keeping the two
cavity frequencies unchanged, while pre-shifting $\omega_\text{cp}$ to
enforce frequency matching. Specifically, by inspection of
\eqreftwo{preshift1}{preshift3}, it follows that choosing input-light
frequencies
\begin{align}
  \w_{0}^{\text{crit}} &= \w_\text{c0} \lp 1 - \alpha_{00}
  |a_0^{\text{crit}}|^2 - \alpha_\text{0m} |a_\text{m}^{\text{crit}}|^2 -
  \alpha_\text{0p} |a_\text{p}^{\text{crit}}|^2 \rp
\label{eq:critshift1} \\
\w_\text{m}^{\text{crit}} &= \w_\text{cm} \lp 1 - \alpha_\text{m0} |a_0^{\text{crit}}|^2 - \alpha_\text{mm} |a_\text{m}^{\text{crit}}|^2 - \alpha_\text{mp} |a_\text{p}^{\text{crit}}|^2 \rp 
\label{eq:critshift2}, 
\end{align}
and tuning $\omega_\text{cp}$ such that
\begin{equation}
 \wpci = { 2 \w_\text{c0} (1 - \sum \a_{0k} |a_k^{\text{crit}}|^2) - \w_\text{cm} (1 - \sum \a_{\text{m}k} |a_k^{\text{crit}}|^2) \over 1 - \sum \a_{\text{p}k} |a_k^{\text{crit}}|^2 } \label{wpcideal},
\end{equation}
yields the same steady-state critical solution above. This approach is
advantageous in that the requirement that all three cavity frequencies
be simultaneously and independently tuned (post-fabrication) is
removed in favor of tuning a single cavity mode.  Given a scheme to
tune the frequencies of the cavity modes that achieves perfect
frequency matching at the critical power, what remains is to analyze
the stability and excitability of the new critical solution, which can
be performed using a staightforward linear stability analysis of the
coupled mode equations~\cite{Drummond80}. Before addressing these
questions, however, is important to address a more serious concern.

\subsection{Frequency mismatch}
\label{sec:mismatch}

Regardless of tuning mechanism, in practice one can never fully
satisfy perfect frequency matching (even when self- and cross-phase
modulation can be neglected) due to fabrication imperfections. In
general, one would expect the finite bandwidth to mean that there is
some tolerance $\sim 1/Q_\text{p}$ on any frequency mismatch $\Delta \omega =
2 \omega_\text{c0}-\omega_\text{cm}-\omega_\text{cp} \lesssim
\omega_\text{cp}/Q_\text{cp}$~\cite{Zhuan12}. However, here we find that
instabilities and strong modifications of the cavity lineshapes
arising from the particular nature of this nonlinear process lead to
extreme, sub-bandwidth sensitivity to frequency deviations that must
be carefully examined if one is to achieve high-efficiency operation.

\begin{figure}[t!]
 \centering
 \includegraphics[width=0.45\textwidth]{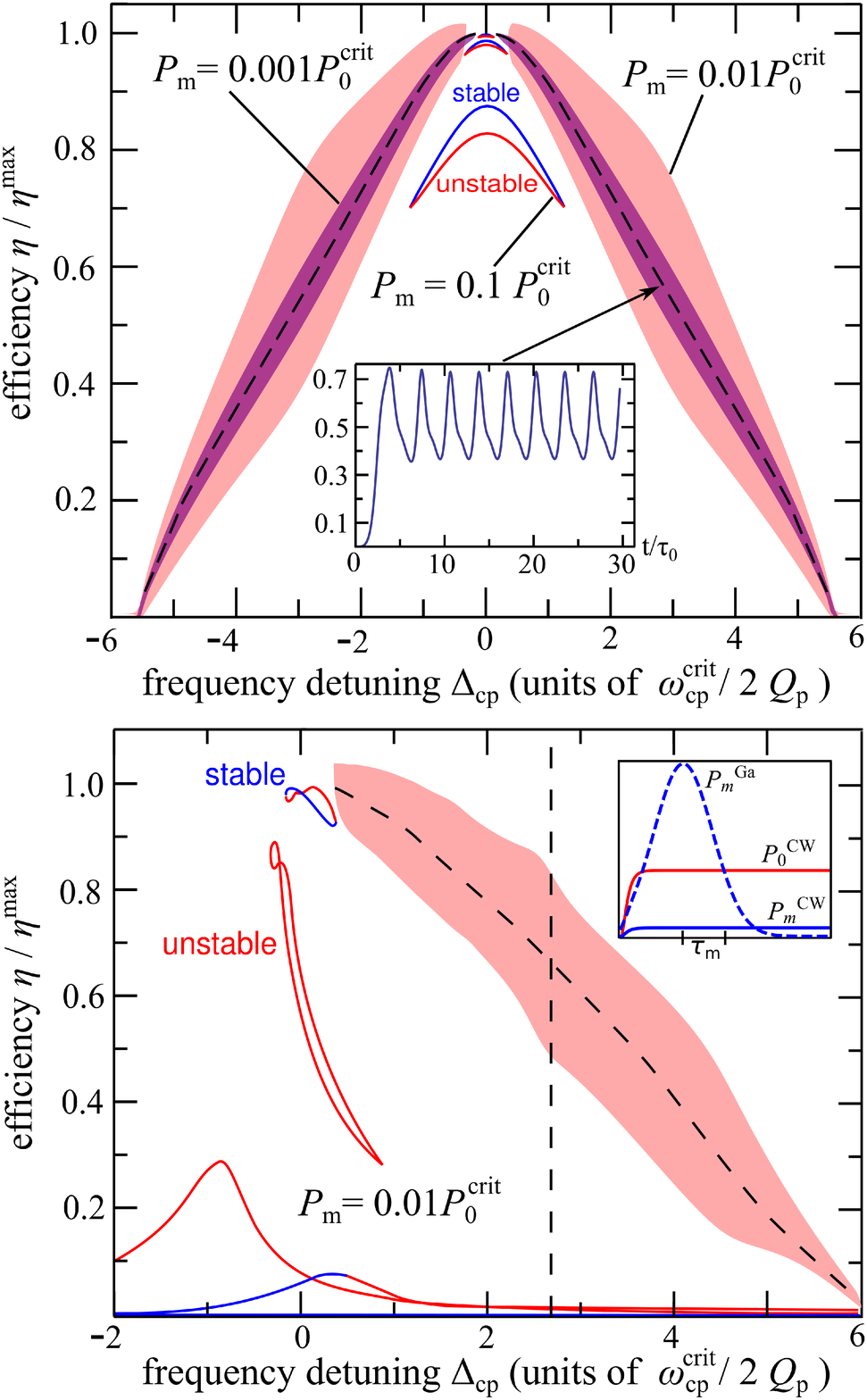}
 \caption{(Top:) Steady-state conversion efficiency $\eta$ (normalized
   by the maximum achievable efficiency $\eta^\text{max}$) as a
   function of frequency mismatch $\Delta_\text{cp} =
   \omega_\text{cp}-\omega_\text{cp}^\textrm{crit}$ (in units of
   $\w_\text{cp}^\text{crit}/2 Q_\text{p}$), for the cavity system depicted in
   \figref{2Dfields}, but in the absence of self- and cross-phase
   modulation ($\alpha=0$). Incident frequencies are chosen to be
   $\w_0= \w_0^\text{crit}$ and $\omega_\text{m} = \w_\text{m}^\text{crit}$, with
   corresponding powers $P_0 = P_0^\text{crit}$ and $P_\text{m}$, where we
   consider multiple $P_\text{m} = \{0.1,~0.01,~0.001\}
   P_0^\text{crit}$. Note that since $\a=0$, critical frequencies are
   independent of incident powers, so that $\w_0^\text{crit} =
   \w_\text{c0}$, $\w_\text{m}^\text{crit} = \w_\text{cm}$, and $\w_\text{cp}^\text{crit} =
   2\w_\text{c0} - \w_\text{cm}$. Blue/red solid lines denote stable/unstable
   fixed points, whereas shaded areas indicate regimes lacking fixed
   point solutions and exhibiting limit-cycle behavior, shown only for
   $P_\text{m} = \{0.01, 0.001\} P_0^\text{crit}$, with smaller amplitudes
   corresponding to smaller $P_\text{m}$. Dashed lines denote the average
   efficiency of the limit cycles $\bar{\eta}$, whereas the top/bottom
   of the shaded regions denote the maximum/minimum efficiency per
   period. The inset shows the efficiency as a function of time for a
   typical limit cycle, obtained at $\Delta_\text{cp} \approx 3
   \w_\text{cp}^\text{crit}/2 Q_\text{p}$.  (Bottom:) $\eta$ and $\bar{\eta}$ for
   the same system above, but in the presence of self- and cross-phase
   modulation ($\alpha \neq 0$), and only for $P_\text{m}=0.01
   P_0^\text{crit}$. Note that additional stable and unstable fixed
   points arise due to the non-zero $\alpha$, and that limit-cycle
   behaviors arise only for $\Delta_\text{cp} > 0$. Inset shows the
   temporal shape of the incident power needed to excite the desired
   limit cycles, corresponding to a Gaussian pulse superimposed over
   CW inputs.}
\label{fig:bwidth}
\end{figure}

To illustrate the effects of frequency mismatch, we first consider an
ideal, lossless system with zero self- and cross-phase modulation
($\alpha=0$) and with incident light at frequencies $\omega_0 =
\omega_\text{c0}$ and $\omega_\text{m} = \omega_\text{cm}$, and powers
$P^\text{crit}_0$ and $P_\text{m}$, respectively. With the exception of
$\alpha$, the coupling coefficients and cavity parameters correspond
to those of the 2d design described in \secref{2d}. \Figref{bwidth}
(top) shows the steady-state conversion efficiency $\eta$ (solid
lines) as a function of the frequency mismatch $\Delta_\text{cp} = \w_\text{cp}
- \wpci$ away from perfect frequency-matching, for multiple values of
$P_\text{m}=\{0.001,0.01,0.1\}P^\text{crit}_0$, with blue/red solid lines
denoting stable/unstable steady-state fixed points. As shown,
solutions come in pairs of stable/unstable fixed points, with the
stable solution approaching the maximum-efficiency $\eta^\text{max}$
critical solution as $P_\text{m} \to 0$. Moreover, one observes that as
$\Delta_\text{cp}$ increases for finite $P_\text{m}$, the stable and unstable
fixed points approach and annihilate one other, with limit cycles
appearing in their stead (an example of what is known as a ``saddle-node
homoclinic bifurcation"~\cite{Shilnikov83}). The mismatch at which this
bifurcation occurs is proportional to $P_\text{m}$, so that, as $P_\text{m} \to 0$,
the regime over which there exist high-efficiency steady states
reduces to a single \emph{fixed point} occurring at $\Delta_\text{cp} =
0$. Beyond this bifurcation point, the system enters a limit-cycle
regime (shaded regions) characterized by periodic modulations of the
output signal in
time~\cite{Strogatz94,Drummond80,Hashemi09}. Interestingly, we find
that the average efficiency of the limit cycles (dashed lines),
\begin{equation}
\bar{\eta} = \lim_{T\to\infty}
  \frac{1}{T} \int_0^T dt\, \eta(t),
\end{equation}
remains large and $\lesssim \eta^\text{max}$ even as $\Delta_\text{cp}$ is
several fractional bandwidths. The inset of \figref{bwidth} (top)
shows the efficiency of this system as a function of time (in units of
the lifetime $\tau_0$) for large mismatch $\Delta_\text{cp} =
3\omega^\text{crit}_\text{cp}/2Q_\text{p}$. As expected, the modulation amplitude
and period of the limit cycles depend on the input power and mismatch,
and in particular we find that the amplitude goes to zero and the
period diverges $\sim 1/\Delta_\text{cp}$ as $\Delta_\text{cp} \to 0$. This
behavior is observed across a wide range of $P_\text{m}$, with larger $P_\text{m}$
leading to lower $\bar{\eta}$ and larger amplitudes. For small enough
mismatch, the modulation frequency enters the THz regime, in which
case standard rectifications procedures~\cite{Tonouchi07} can be
applied to extract the useful THz
oscillations~\cite{Lee00:thz,Morozov05,Vodopyanov06,yeh07,Andronico08,Bravo-AbadRo10}.

Frequency mismatch leads to similar effects for finite $\alpha$,
including homoclinic bifurcations and corresponding high-efficiency
limit cycles that persist even for exceedingly large frequency
mismatch. One important difference, however, is that the redshift
associated with self- and cross-phase modulation creates a strongly
asymmetrical lineshape that prevents high-efficiency operation for
$\Delta_\text{cp} < 0$. \Figref{bwidth} (bottom) shows the stable/unstable
fixed points (solid lines) and limit cycles (dashed regions) as a
function of $\Delta_\text{cp}$ for the same system of \figref{bwidth} (top)
but with finite $\alpha$, for multiple values of
$P_\text{m}=\{0.001,0.01\}P^\text{crit}_0$. As before, the coupling
coefficients and cavity parameters correspond to those of the 2d
design described in \secref{2d}. Here, in contrast to the $\alpha=0$
case, the critical incident frequencies $\omega^\text{crit}_{0}$ and
$\omega^\text{crit}_\text{m}$ are chosen according to
\eqreftwo{critshift1}{critshift2} in order to counter the effects of
self- and cross-phase modulation, and are therefore generally
different from $\omega_\text{c0}$ and $\omega_\text{cm}$. Aside from the
asymmetrical lineshape, one important difference from the $\alpha=0$
case is the presence of additional stable/unstable low-efficiency
solutions. Multistability complicates matters since, depending on the
initial conditions, the system can fall into different stable
solutions and in particular, simply turning on the source at the
critical input power may result in an undesirable low-efficiency
solution. One well-known technique that allows such a system to lock
into the desired high-efficiency solutions is to superimpose a gradual
exponential turn-on of the pump with a Gaussian pulse of larger
amplitude~\cite{Hashemi09}. We found that a single Gaussian pulse with
a peak power of $4 P_0^{\text{crit}}$ and a temporal width $ \sim
\tau_\text{m}$, depicted in the right inset of \figref{bwidth} (bottom), is
sufficient to excite high-efficiency limit cycles in the regime
$\Delta_\text{cp} > 0$.

\begin{figure}[t!]
 \centering
 \includegraphics[width=0.45\textwidth]{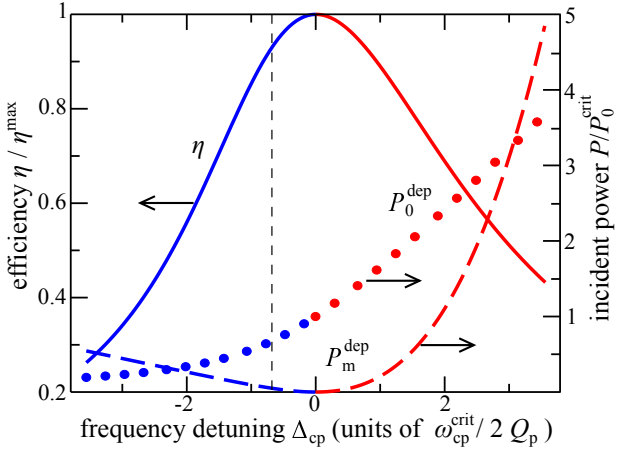}
 \caption{Steady-state conversion efficiency (normalized by the
   maximum achievable efficiency $\eta^\text{max}$) and required
   incident powers $P_0^\text{p}$ and $P_\text{m}^\text{m}$ (normalized by
   the critical power $P_0^\text{crit}$) corresponding to depleted
   steady states of the system of \figref{2Dfields}, as a function of
   frequency mismatch $\Delta_\text{cp} =
   \omega_\text{cp}-\omega_\text{cp}^\textrm{crit}$ (in units of
   $\w_\text{cp}^\text{crit}/2 Q_\text{p}$). As described in \secref{2d-ss},
   depleted states yield 100\% depletion of $P_0$, and are excited by
   appropriate combinations of incident frequencies
   $\w_k=\w_k^\text{dep}$ and powers $P_k = P_k^\text{dep}$.  Blue/red
   lines denote stable/unstable solutions, with solid and dashed
   lines, and circles, denoting $\eta$, $P_\text{m}^\text{dep}$, and
   $P_0^\text{dep}$, respectively.}
\label{fig:2D}
\end{figure}

Despite their high efficiencies (even for large $\Delta_\text{cp}
\gtrsim 1$), the limit-cycle solutions above leave something to be
desired. Depending on the application, it may be desirable to operate
at high-efficiency fixed points. One way to achieve this for non-zero
frequency mismatch is to abandon the critical solution and instead
choose incidence parameters that exploit self- and cross-phase
modulation in order to enforce perfect frequency matching and 100\%
depletion of the pump as follows,
\begin{align}
  \w_\text{cp}^\text{NL} + \w_\text{cm}^\text{NL} &= 2
  \w_\text{c0}^\text{NL}, \label{eq:freqmatch1} \\
  s_{0-} &= 0 \label{eq:freqmatch2}
\end{align}
Specifically, enforcing \eqreftwo{freqmatch1}{freqmatch2} by solving
\eqreftwo{cme1}{cme4} for
$\w_0^\text{dep},~\w_\text{m}^\text{dep},~P_0^\text{dep}$, and
$P_\text{m}^\text{dep}$, we obtain a \emph{depleted} steady-state
solution $a^\text{dep}_k$ that, in contrast to the critical solution
$a^\text{crit}_k$, yields a steady-state efficiency that corresponds
to 100\% depletion of the pump regardless of frequency mismatch. Note
that we are not explicitly maximizing the conversion efficiency but
rather enforcing complete conversion of pump energy in the presence of
frequency mismatch, at the expense of a non-negligible input
$P_m^\text{dep}$. \Figref{2D} shows the depleted steady-state
efficiency $\eta^\text{dep}$ (solid line) and corresponding incident
powers (solid circles and dashed line) as a function of
$\Delta_\text{cp}$, for the same system of \figref{bwidth}
(bottom). We find that for most parameters of interest, depleted
efficiencies and powers are uniquely determined by
\eqreftwo{freqmatch1}{freqmatch2}. As expected, the optimal efficiency
occurs at $\Delta_\text{cp}=0$ and corresponds to the critical
solution, so that $P_0^\text{dep} = P_0^\text{crit}$,
$P_\text{m}^\text{dep}=P_\text{m}^\text{crit}=0$, and $\eta^\text{dep}
= \eta^\text{max}$. For finite $\Delta_\text{cp} \neq 0$, the optimal
efficiencies are lower due to the finite $P_\text{m}^\text{dep}$, but
there exist a broad range of $\Delta_\text{cp}$ over which one obtains
relatively high efficiencies $\sim \eta^\text{max}$. Power
requirements $P^\text{dep}_0$ and $P^\text{dep}_\text{m}$ follow
different trends depending on the sign of $\Delta_\text{cp}$.  Away
from zero detuning, $P_\text{m}^\text{dep}$ can only increase whereas
$P_0^\text{dep}$ decreases for $\Delta_\text{cp}<0$ and increases for
$\Delta_\text{cp}>0$.  In the latter case, the total input power
exceeds $P^\text{crit}_0$ leading to the observed instability of the
fixed-point solutions.

Finally, we point out that limit cycles and depleted steady states
reside in roughly complementary regimes. Although no stable
high-efficiency fixed points can be found in the $\Delta_\text{cp} >
0$ regime, it is nevertheless possible to excite high-efficiency limit
cycles. Conversely, although no such limit cycles exist for
$\Delta_\text{cp}<0$, it is possible in that case to excite
high-efficiency depleted steady states.

\section{Nanobeam designs}
\label{sec:designs}

In this section, we consider concrete and realistic cavity designs in
2d and 3d, and check the predictions of our TCMT by performing exact
nonlinear FDTD simulations in 2d. Our designs are based on a
particular class of PhC nanobeam structures, depicted schematically in
\figreftwo{2Dfields}{3Dfields}, where a cavity is formed by the
introduction of a defect in a lattice of air holes in dielectric, and
coupled to an adjacent waveguide formed by the removal of holes on one
side of the defect. We restrict our analysis to dielectric materials
with high nonlinearities at near- and mid-infrared
wavelengths~\cite{Boyd92}, and in particular focus on undoped silicon,
whose refractive index $n \approx 3.4$ and Kerr susceptibility $\ch
\sim 10^{-18}~m^2/V^2$~\cite{Lin07}.

\subsection{General design considerations}
\label{sec:general}

Before delving into the details of any particular design, we first
describe the basic considerations required to achieve the desired high
efficiency characteristics. To begin with, we require three modes
satisfying the frequency-matching condition to within some desired
bandwidth (determined by the smallest of the mode bandwidths). We
begin with the linear cavity design, in which case we seek modes that
approximately satisfy $\w_\text{cm} + \w_\text{cp} = 2 \w_\text{c0}$. The final
cavity design, incorporating self- and cross-phase modulation, is then
obtained by additional tuning of the mode frequencies as described
above.  Second, we seek modes that have large nonlinear overlap
$\beta$, determined by \eqref{beta}. (Ideally, one would also optimize
the cavity design to reduce $\alpha/\beta$, but such an approach falls
beyond the scope of this work.) Note that the overlap integral $\beta$
replaces the standard ``quasiphase matching'' requirement in favor of
constraints imposed by the symmetries of the cavity~\cite{Boyd92}. In
our case, the presence of reflection symmetries means that the modes
can be classified as either even or odd and also as ``TE-like''
($\vec{E} \cdot \hat{\vec{z}}\approx 0$) or ``TM-like'' ($\vec{H}
\cdot \hat{\vec{z}}\approx 0$)~\cite{JoannopoulosJo08-book}, and hence
only certain combinations of modes will yield non-zero overlap.  It
follows from \eqref{beta} that any combination of even/odd modes will
yield non-zero overlap so long as $\vE{\text{m}}$ and $\vE{\text{p}}$ have the same
parity, and as long as all three modes have similar polarizations:
modes with different polarization will cause the term $\sim \lp
\vec{E}_0^* \cdot \vec{E}_\text{m} \rp \lp \vec{E}_0^* \cdot \vec{E}_\text{p} \rp$
in \eqref{beta} to vanish.  Third, in order to minimize radiation
losses, we seek modes whose radiation lifetimes are much greater than
their total lifetimes, as determined by any desired operational
bandwidth. In what follows, we assume operational bandwidths with $Q
\sim 10^3$. Finally, we require that our system support a single
input/output port for light to couple in/out of the cavity, with
coupling lifetimes $Q_{\text{s}k} \ll Q_{rk}$ in order to have negligible
radiation losses.

\subsection{2d design}
\label{sec:2d}

\begin{figure}[t!]
 \centering
 \includegraphics[width=0.45\textwidth]{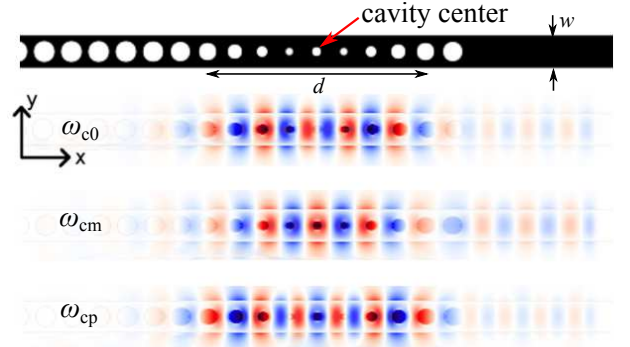}
 \caption{Schematic of two-dimensional, triply-resonant cavity design
   involving a PhC nanobeam of refractive index $n=3.4$, width
   $w=1.2a$ and adiabatically varying hole radii (see text). The
   effective cavity length $d = 6.6a$ and the radius of the central
   hole $R_0$ are chosen so as to fine-tune the relative frequency
   spacing and lifetimes of the modes.  Also shown are the $E_y$
   electric field components of the three modes relevant to DFWM. The
   cavity is coupled to a waveguide formed by the removal of holes to
   the right of the defect.}
   \label{fig:2Dfields}
\end{figure}

In what follows, we consider two different 2d cavities with different
mode frequencies but similar lifetimes and coupling
coefficients. (Note that by 2d we mean that electromagnetic fields are
taken to be uniform in the $z$ direction.)  The two cavities follow the same
backbone design shown in~\figref{2Dfields} which supports three
TE-polarized modes ($\vec{H}\cdot \vec{\hat{z}}=0$) with radiative
lifetimes $Q^{\text{rad}}_0 = 6 \times 10^4,~Q^{\text{rad}}_\text{m} = 6
\times10^4,$ and $Q^{\text{rad}}_\text{p} = 3 \times 10^3$, and total
lifetimes $Q_0 = 1200,~Q_\text{m} = 1100,$ and $Q_\text{p} = 700$, respectively. The
nonlinear coupling coefficients are calculated from the linear modal
profiles (shown on the inset of \figref{2Dfields}) via
\eqreftwo{beta}{nlcoef4}, and are given by: \small
\begin{equation}
\beta = (23.69 + 5.84 i)\times 10^{-5} \ablunit, \notag 
\end{equation}
\begin{align}
  \a_{00} &= 4.935\times10^{-4} \ablunit,~ \a_\text{mm} = 5.096\times10^{-4} \ablunit,& \notag \\
  \a_\text{pp} &= 4.593\times10^{-4} \ablunit,~ \a_\text{0m} = 6.540\times10^{-4} \ablunit,& \notag \\
  \a_\text{0p} &= 5.704\times10^{-4} \ablunit,~ \a_\text{mp} =
  5.616\times10^{-4} \ablunit,& \notag
\end{align}
\normalsize where the additional factor of $h$ allows comparison to
the realistic 3d structure below and accounts for finite nanobeam
thickness (again, assuming uniform fields in the $z$
direction). Compared to the optimal $\beta^{max} = {3 \over 4 n^4 w d
} \lp\frac{\ch}{\epsilon_0 h}\rp$, corresponding to modes with uniform
fields inside and zero fields outside the cavity, we find that $\beta
= 5.5 \times 10^{-3} \beta^{max}$ is significantly smaller due to the
fact that these TE modes are largely concentrated in air. In the 3d
design section below, we choose modes with peaks in the dielectric
regions, which leads to much larger $\beta \approx 0.4\beta^{max}$.

\begin{figure}[t!]
 \centering
 \includegraphics[width=0.45\textwidth]{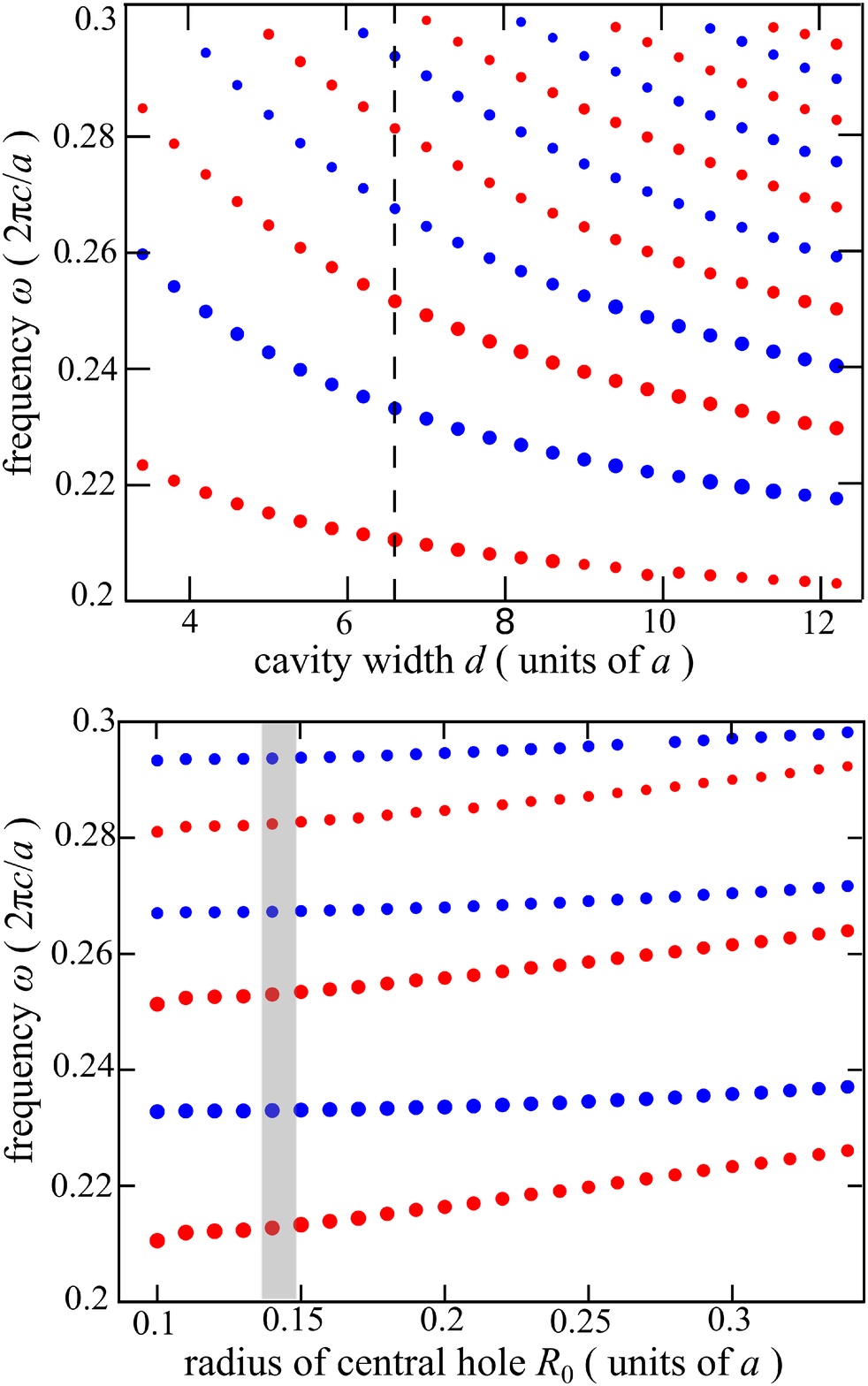}
 \caption{Mode frequencies (in units of $2\pi c/a$) as a function of
   effective cavity length $d$ (top), for fixed center-hole radius
   $R_0 = 0.9a$, and as a function of $R_0$ (bottom), for fixed
   $d=6.6a$. Red/blue circles indicate symmetric/anti-symmetric mode
   profiles, where the size of the circle is proportional to the modal
   lifetime (quality factor) of the corresponding mode. The shaded
   area indicates the parameter region explored in the Sec. B1, B2.}
   \label{fig:tuning}
\end{figure}

In order to arrive at this 2d design, we explored a wide range of
defect parameters, with the defect formed by modifying the radii of a
finite set of holes in an otherwise periodic lattice of air holes of
period $a$ and radius $R=0.36a$ in a dielectric nanobeam of width
$w=1.2a$ and index of refraction $n=3.4$. The defect was parametrized
via an exponential adiabatic taper of the air-hole radii $r$, in
accordance with the formula $r(x)=R \lp 1 - {3 \over 4} e^{-{4\lfloor
    x \rfloor^2 \over d^2}} \rp$, where the parameter $d$ is an
``effective cavity length''. Such an adiabatic taper is chosen to
reduce radiation/scattering losses at the interfaces of the
cavity~\cite{Palamaru01}. The removal of holes on one side of the
defect creates a waveguide, with corresponding cavity--waveguide
coupling lifetimes $Q_{\text{s}k}$ determined by the number of holes
removed~\cite{Kim04,Faraon07,Banaee07}. To illustrate the dependence
of the mode properties on the cavity parameters, \figref{tuning}(top)
shows the evolution of the cavity-mode frequencies as a function of
$d$, with blue/red dots denoting even/odd modes and with larger dots
denoting longer modal lifetimes. As expected, the volumes of the modes
decrease with decreasing $d$, leading to larger $\beta$ (smaller
critical powers) but causing the frequency gap between the modes and
radiation losses to increase. We find that the desired modal
parameters for FWM lie at some intermediate $d \approx 6.6a$. In order
to tune the relative frequencies between the modes, an additional
tuning parameter is required. Specifically, it follows from
perturbation theory~\cite{Joannopoulos95} that changing the central
hole radius $R_0$ allows control of the even-mode frequencies while
leaving odd-mode frequencies unchanged. \Figref{tuning}(bottom) shows
the evolution of the cavity-mode frequencies as a function of $R_0$,
for a fixed $d=6.6a$. As described below, the particular choice of
$R_0$ will depend on whether one seeks to operate with high-efficiency
limit cycles versus high-efficiency steady-state solutions.

\subsubsection{Limit cycles}
\label{sec:2d-lim}

\begin{figure}[t!]
 \centering
 \includegraphics[width=0.45\textwidth]{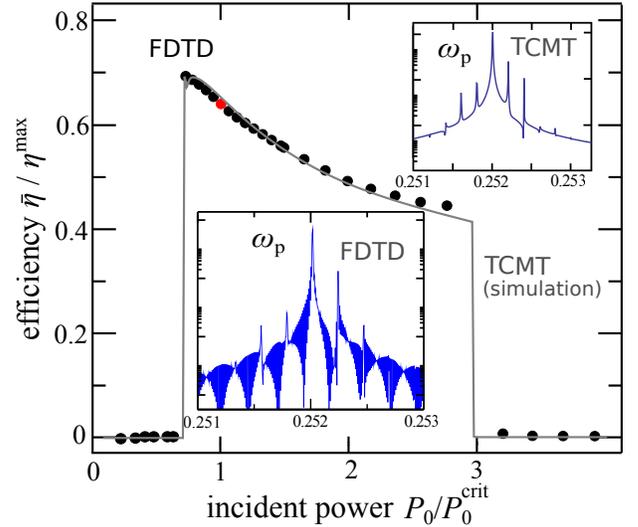}
 \caption{Average conversion efficiency $\bar{\eta}$ (normalized by
   the maximum achievable efficiency $\eta^\text{max}$) of limit
   cycles as a function of power $P_0$ (normalized by
   $P_0^\text{crit}$) at the critical frequencies
   $\omega_0^\text{crit}$ and $\omega_\text{m}^\text{crit}$ and a
   fixed $P_\text{m} = 0.01 P_0^\text{crit}$.  The modal parameters
   are obtained from the 2d cavity of \figref{2Dfields}, with chosen
   $R_0=0.149a$ leading to a detuning $\Delta_\text{cp} \approx
   3\omega_\text{cp}^\text{crit}/2Q_\text{p}$ corresponding to the
   dashed line in~\figref{bwidth} (bottom). Solid circles and gray
   lines denote results as computed by FDTD and TCMT. Insets show the
   spectra of the output light for a given $P_0$ (red circle), and for
   both FDTD and TCMT.}
 \label{fig:lcfit}
\end{figure}

\begin{figure}[t!]
 \centering
 \includegraphics[width=0.45\textwidth]{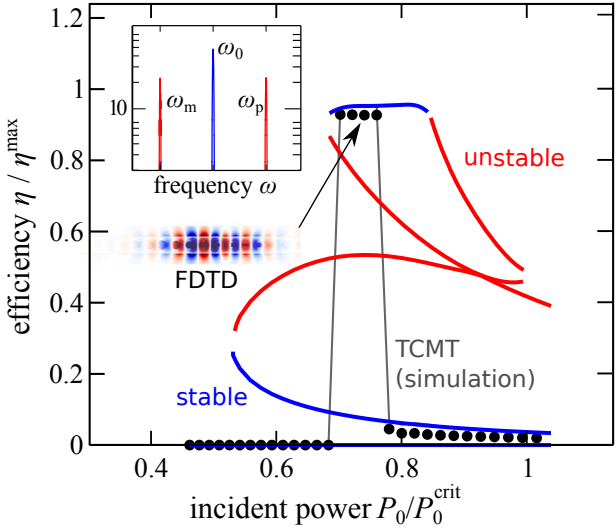}
 \caption{Conversion efficiency $\eta$ (normalized by the maximum
   achievable efficiency $\eta^\text{max}$) of depleted states as a
   function of power $P_0$ (normalized by $P_0^\text{crit}$), at
   incident frequencies $\w_0^\text{dep}$ and
   $\w_\text{m}^\text{dep}$, and a fixed power $P_\text{m}^\text{dep}
   \approx 0.2 P_0^\text{crit}$. The modal parameters are obtained
   from the 2d cavity of \figref{2Dfields}, with $R_0=0.143 a$ leading
   to a detuning $\Delta_\text{cp} \approx
   -0.6\omega_\text{cp}^\text{crit}/2Q_\text{p}$ corresponding to the
   dashed line in~\figref{2D}.  $E_y$ component of the steady-state
   electric field inside the cavity is shown as an inset
   (left-bottom).  Solid circles and gray lines denote FDTD and TCMT,
   while blue/red lines denote stable/unstable steady states. Inset
   (left-top) shows the spectral profile (in arbitrary units) of the
   system, showing full depletion of the pump (blue) and
   correspondingly high conversion of the signal/idler frequencies
   (red). For $P_0 \gtrsim 0.8 P_0^\text{crit}$, the system becomes
   ultra sensitive to the priming parameters, in which case
   high-efficiency solutions can only be excited by adiabatic tuning
   of the pump power (see text).}
\label{fig:MTfit}
\end{figure}

In this section, we consider a design supporting high-efficiency limit
cycles. Choosing $R_0 = 0.149a$, we obtain critical parameters
$\w_0^{\text{crit}} = 0.2319 \wunit $, $\w_\text{m}^{\text{crit}} = 0.2121
\wunit $, $\w_\text{cp}^{\text{crit}} = 0.2530 \wunit$, and
$P^{\text{crit}}_0 = 10^{-3} ({2 \pi c \epsilon_0 a h \over
  \chi^{(3)}})$, corresponding to frequency mismatch $\Delta_\text{cp}
\approx 3\omega_\text{cp}^\text{crit}/2Q_\text{p}$ and critical efficiency
$\eta^\text{max}=0.51$. Choosing a small but finite $P_\text{m} = 0.01
P_0^\text{crit}$, it follows from \figref{bwidth} (dashed line) that
the system will support limit cycles with average efficiencies
$\bar{\eta} \approx 0.65 \eta^\text{max}$. To excite these solutions,
we employed the priming technique described in
\secref{mismatch}. \Figref{lcfit} shows $\bar{\eta}$ as a function of
$P_0$, for incident frequencies $\omega_{k} = \omega^\text{crit}_k$
determined by \eqreftwo{critshift1}{critshift2}, as computed by our
TCMT (gray line) and by exact, nonlinear FDTD simulations (solid
circles).  The two show excellent agreement.  For $0.7 <
P_0/P_0^\text{crit} <3$, we observe limit cycles with relatively high
$\bar{\eta}$ , in accordance with the TCMT predictions, whereas
outside of this regime, we find that the system invariably falls into
low-efficiency fixed points. The periodic modulation of the limit
cycles means that instead of a single peak, the spectrum of the output
signal consists of a set of equally spaced peaks surrounding
$\omega_\text{p}$. The top and bottom insets of \figref{lcfit} show the
corresponding frequency spectra of the TCMT and FDTD output signals
around $\omega_{p}$, for a particular choice of $P_0 \approx
P_0^\text{crit}$ (red circle), showing agreement both in the relative
magnitude and spacing $\approx 2.5\times 10^{-3} \wunit$ of the peaks.

\subsubsection{Depleted steady states}
\label{sec:2d-ss}

In this section, we consider a design supporting high-efficiency,
depleted steady states. Choosing $R_0 = 0.143a$, one obtains critical
parameters $\w_0^{\text{crit}} = 0.2320 \wunit $, $\w_\text{m}^{\text{crit}}
= 0.2118 \wunit $, $\w_\text{cp}^{\text{crit}} = 0.2532 \wunit$, and
$P^{\text{crit}}_0 = 10^{-3} ({2 \pi c \epsilon_0 a h \over
  \chi^{(3)}})$, corresponding to frequency mismatch $\Delta_\text{cp}
\approx -0.6\omega_\text{cp}^\text{crit}/2Q_\text{p}$ and critical efficiency
$\eta^\text{max}=0.51$. Choosing incident frequencies
$\ws{0}^\text{dep}= 0.2320 \wunit$, $\ws{\text{m}}^\text{dep}=0.2119 \wunit$,
and incident powers $P_0^\text{dep} \approx 0.7 \pcrit$ and
$P_\text{m}^\text{dep} \approx 0.04 \pcrit$, it follows from \figref{2D}
(dashed line) that the system supports stable, depleted steady states
with efficiencies $\approx 0.95 \eta^\text{max}$.  \Figref{MTfit}
shows the efficiency of the system as a function $P_0$, with all other
incident parameters fixed to the depleted-solution values above, where
blue/red lines denote stable/unstable solutions. As before, we employ
the priming technique of \secref{mismatch} in order to excite the
desired high efficiency solutions and obtain excellent agreement
between our TCMT (gray line) and FDTD simulations (solid circles). 
Exciting the high-efficiency solutions by steady-state input ``primed" 
with a Gaussian pulse is convenient in FDTD because it leads to 
relatively short simulations, but is problematic for $P_0>0.8P_0^\text{crit}$, 
where the system becomes very sensitive to the priming parameters, 
and it became impractical in for us to find the optimal FDTD source 
conditions in~\figref{MTfit}. In realistic experimental situations, 
however, one can use a different technique to excite the high-efficiency 
solution in a way that is very robust to errors, based on adiabatic tuning of the pump power~\cite{Hashemi09}.


\subsection{3d design}
\label{sec:3d}

\begin{figure}[t!]
 \centering
 \includegraphics[width=0.48\textwidth]{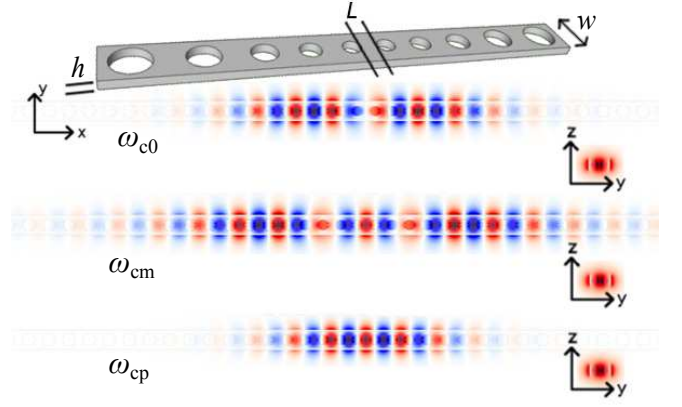}
 \caption{Schematic of three dimensional, triply resonant cavity
   design involving a PhC nanobeam of refractive index $n=3.4$, width
   $w=a$, and height $h=0.51a$, and linearly tapered air holes, as
   described in the text. The central cavity length $L \approx 0.4a$
   and number of taper segments are chosen so as to fine-tune the
   relative frequency spacing and lifetimes of the modes. Also shown
   are the $E_y$ electric-field components of three TE-like modes with
   fundamental TE00 transverse profiles, and with frequencies
   $\w_\text{c0}=0.2848 \wunit,~ \w_\text{cm} = 0.2801 \wunit$ and $\w_\text{cp} =
   0.2895 \wunit$.  Radiation lifetimes are found to be
   $Q^{\text{rad}}_0 = 10^6,~Q^{\text{rad}}_\text{m}=3\times 10^4$, and
   $Q^{\text{rad}}_\text{p} = 2 \times 10^4$.}
\label{fig:3Dfields}
\end{figure}

We now consider a 3d design, depicted in \figref{3Dfields}, as a
feasible candidate for experimental realization. The cavity supports
three TE$_{00}$ modes ($\vec{E}_z = 0 \text{ at } z=0$) of frequencies
$\w_\text{c0} = 0.2848 \wunit,~ \w_\text{cm} = 0.2801 \wunit$ and $\w_\text{cp} =
0.2895 \wunit$, radiative lifetimes $Q^{\text{rad}}_0 =
10^6,~Q^{\text{rad}}_\text{m}=3\times 10^4,~Q^{\text{rad}}_\text{p} = 2 \times
10^4$. As before, the total lifetimes can be adjusted by removing air
holes to the right or left of the defect, which would allow coupling
to the resulting in-plane waveguides. (Alternatively, one might
consider an out-of-plane coupling mechanism in which a fiber carrying
incident light at both $\omega_0$ and/or $\omega_\text{m}$ is brought in
close proximity to the cavity~\cite{Kim04,Barclay05}.) In what
follows, we do not consider any one particular coupling channel and
focus instead on the isolated cavity design. Nonlinear coupling
coefficients are calculated from the linear modal profiles (shown on
the inset of \figref{3Dfields}) via \eqreftwo{nlcoef1}{nlcoef4}, and
are given by: \small
\begin{equation}
\beta = 2 \times 10^{-4} \abunit, \notag 
\end{equation}
\begin{align}
\a_{00} &= 8.1 \times10^{-4} \abunit,~ \a_\text{mm} = 4.6 \times10^{-4} \abunit,& \notag \\
\a_\text{pp} &= 11.5 \times10^{-4} \abunit,~ \a_\text{0m} = 6.2 \times10^{-4} \abunit,& \notag \\
\a_\text{0p} &= 12.7 \times10^{-4} \abunit,~ \a_\text{mp} = 5.5 \times10^{-4} \abunit. \notag
\end{align}
Here, in contrast to the 2d design of \secref{2d}, we chose modes
whose amplitudes are concentrated in dielectric regions, and therefore
find appreciably larger$\beta \approx 0.4 \beta^{max}$.

In order to arrive at the above 3d design, we explored a cavity
parametrization similar to the one described
in~\cite{Parag09}. Specifically, we employed a suspended nanobeam of
width $w = a$, thickness $h = 0.51 a$, and refractive index
$n=3.4$. The beam is schematically divided into a set of $2N$ lattice
segments, each having length $a_i, i \in \{\pm 1, ... ,\pm N\}$ and
corresponding air-hole radii $R_i = 0.3 a_i$, where $a_1$ ($a_{-1}$)
is the length of the lattice segment immediately to the right (left)
of the beam's center. The cavity defect is induced via a linear taper
of $a_i$ over a chosen set of $2\bar{N}$ segments, according to the
formula:
\begin{align}
a_i &=  a \lp f_a + {(1-f_a) \over (\bar{N}-1) }(|i|-1) \rp,&\quad &|i| \leq \bar{N} \notag \\
    &= a ,&\quad &|i|> \bar{N}. \notag
\end{align} 
In order to arrive at our particular design, we chose
$f_a=0.85,~N=21,~\bar{N}=9$ and varied the central cavity length $L$
to obtain the desired TE$_{00}$ modes. Assuming total modal lifetimes $Q_0 =
8500,~Q_\text{m} = 3000$, and $Q_\text{p} = 3000$ and using these design parameters,
we obtain critical parameters $\w_0^{\text{crit}} = 0.2843 \wunit$,
$\w_\text{m}^{\text{crit}} = 0.2798 \wunit$, $\wpci = 0.2895 \wunit$, and
$P_0^{\text{crit}} = 5\times10^{-5} \punit$, corresponding to
frequency mismatch $\Delta_\text{cp} \approx -0.07 \wpci/(2 Q_\text{p})$ and
$\eta^\text{max}= 0.42$. Note that because the radiative losses in
this system are non-negligible, the maximum efficiency of this system
is $\approx 82\%$ of the optimal achievable efficiency $\w_\text{p}/(2 \w_0)
\approx 0.51$. At these small $\Delta_\text{cp}$, we find that depletion of
the pump is readily achieved through the critical parameters
associated with perfect frequency matching. However, as illustrated in
\secref{2d-lim}, it is indeed possible to choose a design that leads
to highly efficient limit cycles or other dynamical behaviors.

We now express the power requirements of this particular design using
real units instead of the dimensionless units of $2\pi c\varepsilon_0
a^2 / \ch$ we have employed thus far. Choosing to operate at telecom
wavelengths $\lambda_\text{c0} \equiv 2\pi c/\omega_\text{c0} =
1.5\mu$m, with corresponding $n\approx 3.4$ and $\ch = 2.8 \times
10^{-18} {m^2 \over V^2}$ \cite{Boyd92}, we find that $a = 0.2848
\times 1500 = 427$nm and $P^\text{crit}_0 \approx
50\mathrm{m}W$. Although our analysis above incorporates effects
arising from linear losses (e.g. due to material absorption or
radiation), it neglects important and detrimental sources of nonlinear
losses in the telecom range, including two-photon and free carrier
absorption~\cite{Liang04,CCW07}. Techniques that mitigate the latter
exist, e.g. reverse biasing~\cite{Rong06}, but in their absence it may
be safer to operate in the spectral region below the half-bandgap of
silicon~\cite{Lin07}. One possibility is to operate at
$\lambda_\text{c0} = 2.2\mu$m, in which case $\ch \approx 1.5 \times
10^{-18} {m^2 \over V^2}$~\cite{Lin07}, leading to $a = 627$nm and
$P^\text{crit}_0 \approx 200\mathrm{m}W$. For a more detailed analysis
of nonlinear absorption in triply resonant systems, the reader is
referred to~\citeasnoun{Zeng13}. While that work does not consider the
effects of nonlinear dispersion, self- and cross-phase modulation, or
frequency mismatch, it does provide upper bounds on the maximum
efficiency in the presence of two-photon and free-carrier absorption.

\section{Concluding remarks}

In conclusion, using a combination of TCMT and FDTD simulations, we
have demonstrated the possibility of achieving highly efficient DFWM
at low input powers ($\sim 50\mathrm{m}W$) and large bandwidths ($Q
\sim 1000$) in a realistic and chip-scale ($\mu$m) nanophotonic
platform consisting of a triply resonant silicon nanobeam cavity. Our
theoretical analysis includes detrimental effects stemming from linear
losses, self- and cross-phase modulation, and mismatch of the cavity
mode frequencies (e.g. arising from fabrication imperfections), and is
checked against the predictions of a full nonlinear Maxwell FDTD
simulation. Although power requirements in the tens of m$W$s are not
often encountered in conventional chip-scale silicon nanophotonics,
they are comparable if not smaller than those employed in conventional
centimeter-scale DFWM schemes~\cite{Ong13,Rong06,Yamada06}. Our
proof-of-concept design demonstrates that full cavity-based DFWM not
only reduces device dimensions down to $\mu$m scales, but also allows
depletion of the pump with efficiencies close to unity. However, we
emphasize that there is considerable room for additional design
optimization. In particular, we find that increasing the radiative
lifetimes of the signal and converted modes (currently almost two
orders of magnitudes lower than the pump) can significantly lower the
power requirements of the system. 

{\it Acknowledgements:} We acknowledge support from the MIT
Undergraduate Research Opportunities Program and the U.S. Army
Research Office through the Institute for Soldier Nanotechnology under
contract W911NF-13-D-0001.

\bibliography{photon}

\end{document}